\documentclass{IEEEtran}
\usepackage{cite}
\ifCLASSINFOpdf
\else
\fi

\usepackage[cmex10]{amsmath}
\usepackage{array}
\usepackage{mdwmath}
\usepackage{mdwtab}
\usepackage{graphicx}
\usepackage{euscript}
\usepackage{amsfonts}
\usepackage{bigstrut}
\usepackage{tabularx}
\usepackage{cite}
\usepackage{amsmath}
\usepackage{algorithm}
\usepackage[noend]{algpseudocode}
\makeatletter
\def\BState{\State\hskip-\ALG@thistlm}
\makeatother
\usepackage{enumitem}

\usepackage{amsmath}
\usepackage{amsthm}
\usepackage{amssymb}
\usepackage{xcolor}

\usepackage{hyperref}

\usepackage{epstopdf} 

\usepackage{bbm}

\DeclareMathOperator*{\argmin}{arg\,min}

\begin{document}

\title{On the Bound of Energy Consumption in Cellular IoT Networks}
\author{Bassel Al Homssi,~\IEEEmembership{Student Member,~IEEE}, Akram~Al-Hourani,~\IEEEmembership{Senior Member,~IEEE},\\Sathyanarayanan Chandrasekharan~\IEEEmembership{Member,~IEEE}, Karina Mabell Gomez,~\IEEEmembership{Member,~IEEE},\\ and Sithamparanathan Kandeepan,~\IEEEmembership{Senior Member,~IEEE}.
\thanks{B. Al Homssi, A. Al-Hourani, S. Chandrasekharan, K. M. Gomez, and S. Kandeepan are with the School of Engineering, RMIT University, Melbourne, Australia. E-mail: akram.hourani@rmit.edu.au}
}

\markboth{IoT Energy}%
{Shell \MakeLowercase{\textit{et al.}}: Bare Demo of IEEEtran.cls for Journals}

\maketitle

\begin{abstract}
Billions of sensors are expected to be connected to the Internet through the emerging Internet of Things (IoT) technologies. Many of these sensors will primarily be connected using wireless technologies powered using batteries as their sole energy source which makes it paramount to optimize their energy consumption. In this paper, we provide an analytic framework of the energy-consumption profile and its lower bound for an IoT end device formulated based on Shannon capacity. We extend the study to model the average energy-consumption performance based on the random geometric distribution of IoT gateways by utilizing tools from stochastic geometry and real measurements of interference in the ISM-band. Experimental data, interference measurements and Monte-Carlo simulations are presented to validate the plausibility of the proposed analytic framework, where results demonstrate that the current network infrastructures performance is bounded between two extreme geometric models. This study considers interference seen by a gateway regardless of its source.
\end{abstract}


\IEEEpeerreviewmaketitle 

\section{Introduction}
\IEEEPARstart{T}{he} overwhelming need for a better-connected world is urging technologists, regulators, and researchers to continuously devise better solutions for coping with the accelerating ICT demand. A projected estimation anticipates a colossal increase in the use of IoT devices, predicting that by 2022, 18 billion IoT devices will join the global network \cite{mobility_report}. This vast growth is driven by many emerging business cases such as utility metering, precision agriculture, vending machines among many others. As promising as it might seem, the current capability of the cellular network is insufficient in handling the massive number of ultra low power IoT devices. The current wireless networks are designed for high speed communication with high Quality of Service (QoS) while using complex access and spectrum utilization methods to cater for efficient multimedia delivery, whereas IoT will rely on low QoS data delivery with simple access to simplify end-devices such as limited-capability sensors. A big limitation arises in the form of energy constraints for IoT devices, where many of the envisioned applications are solely reliant on battery power such as (i) gas and water metering, (ii) monitoring sensors for agriculture and environment, and (iii) tracking of assets. 

In order to tackle IoT specific needs and cater for the current limitations in the cellular network, multiple IoT solutions and standards have been proposed. These standards vary in their solution approach, posing a trade-off challenge on the engineering and business formulation. Such trade-offs include the data (packet size), data throughput, interference mitigation capability, reliability, delay, and energy efficiency \cite{7815384}. In terms of network architecture; some technologies allow mesh-type topologies for lowering infrastructure costs, while others like the current cellular network are based solely on star topologies.

In this paper, we explore the theoretical bounds of energy-consumption of IoT end-devices. In particular, we utilize tools from stochastic geometry to study the effect of network regularity\cite{6841639} on the energy-consumption of end IoT devices in a congested cellular network. Utilizing the well-established Shannon model, we derive practical lower energy consumption bounds in a concise analytical form to provide a useful benchmark to design engineers and researchers alike. The contributions of this paper can be summarized as the following:
\begin{itemize}
\item Derivation of an analytical lower bound for energy consumption in IoT networks, to the authors knowledge such a lower bound does not exist in the literature.
\item Derivation of the optimal (minimum) energy consumption of IoT systems based on the theoretical analysis.
\item The analytical framework of using stochastic geometry for the energy consumption analysis, in particular for IoT systems.
\item Validation of the theoretical lower bounds by means of Monte-Carlo simulations.
\item Application of the lower bound and optimization of energy consumption to a practical scenario, and hence the validation of our proposed work on practical systems.
\end{itemize}
We make the assumption that the IoT network is highly congested with millions of IoT devices sharing the frequency band based on out measurements in Melbourne \cite{8403749}. An example of a congested network is projected to exist in the UK to cater for smart water metering~\cite{LIBERG20181}. In this work we do not include theoretical interference calculations of the network but consider real measured interference in that band. This approach is suitable for frequency bands such as the ISM-band where a lack of coordination between users and networks that coexist leads to minimum control over the interference levels unlike conventional cellular networks where telecom operators are in control of the power, traffic, and access in the frequency band. Note that in this study, we consider interference power statistics on the frequency band of interest based on~\cite{8403749}, while the distributions of the gateways affect the network regularity only and does not contribute to interference statistics.

The rest of this paper is organized as follows. Section~\ref{Sec_Back} lists the relative work to this paper that has been presented in the literature while Section~\ref{Sec_EnerMod} presents an energy model based on the famous Shannon channel capacity model. The model is further analyzed in Section~\ref{Sec_EnerBound} in terms of the restrictions and theoretical bounds in terms of the SINR. Section~\ref{Sec_Cellular} sheds light on the behavior of the energy-consumption in an IoT cellular network. Section~\ref{Sec_Operator} presents a comparison on the energy-consumption boundaries between theoretical network models and a currently deployed network. Finally, Section~\ref{Sec_Conc} provides conclusion remarks and future work.

\section{Background and Related Work}\label{Sec_Back}
Unlike current cellular devices, IoT devices have energy sources that have limited capabilities such as a conventional battery to keep the device simple and mobile. Various Low Power Wide Area Networks (LPWANs) have been proposed to address the energy-consumption problem. This includes the latest release of NB-IoT adopted by many telecommunication operators and supported by the third generation partnership project (3GPP)~\cite{7876968}, Wi-SUN backed by Wi-SUN alliance in-line with IEEE 802.15.4g standard~\cite{6190698}, Sigfox~\cite{7925650}, and also the widely deployed LoRaWAN supported by LoRa Alliance~\cite{8036410}. As a result, comparisons between those technologies have been made in the literature to explore their strengths and limitations and we invite the reader to go through them~\cite{7815384, 7894201,7962714}.

The regularity of a network is a measure of how well located the gateways are for a deterministic number of points~\cite{1649111}. To quantify the network regularity, stochastic geometry is one of the primary tools used to model cellular networks' behavior mathematically. Most of the literature that tackles cellular networks focus on the interference distribution and outage probability in a network or collection of networks in the downlink performance~\cite{6576422,8023448,8368129} which are mostly applicable to technologies operated by telecos (CIoT) such as; NB-IoT, EC-GSM-IoT, and LTE-M which utilize licensed frequency bands. Nevertheless, some recent papers emerged studying the uplink performance in cellular networks~\cite{Elsawy2014, 7155510, Novlan2013} but also can be applied to CIoT technologies for the same reason. Other papers proposed interference models for a single access point~\cite{7803607, Lim2018, 8430542} using LoRaWAN technology in class-licensed frequency bands and cannot be generalized to multi access point scenarios.

On the other hand, this paper proposes a methodology that is technology agnostic and utilizes measured interference in class-licensed frequency bands. It is worthy to note that the class-licensed spectrum is not exclusive to IoT devices \cite{ACMA_LIPD}. In addition, the paper models the lower bound of the energy consumption and quantifies the effect of network regularity on the electric energy consumption, thus to the authors' knowledge these contributions have not been previously addressed in the literature.

\section{IoT Device Energy-Consumption Model}\label{Sec_EnerMod}
A typical IoT end-device mainly consists of three entities; (i) a sensing entity that acquires the external conditions such as temperature, humidity, water-flow meter, etc.., (ii) a processor that handles the housekeeping activities and interfaces to both the sensor and the communication module, and (iii) a communication module that transmits the acquired data to the network. Our interest in this paper is focused on the communication module, particularly on the energy-consumption levels during the uplink transmission. The downlink window is typically used for acknowledgments or updates and is not as frequently used. Moreover, it is envisioned that most of the ultra-low end-devices will mainly be utilized for uplink communications~\cite{specification2015lora}. Assuming that typical uplink messages in the network consist of a concise length $N_m$ in bits including the protocol overhead, the shortest time that the message would take to be transmitted depends on the channel capacity which can be formulated based on Shannon model as,
\begin{table}
	\caption{Notations and Symbols}
	\centering
	\begin{tabularx}{3.49in}{l l l}
		\hline\hline \\[-1.5ex]
		Symbol & Value &Definition \\
		\hline\\ [-0.5ex]
		$N_m$ &144~bits &Number of bits in a message. \\
		$T_m$ &- &Message time (time-on-air). \\
		$B$ &125~kHz &Channel bandwidth \\
		$\gamma$ &- &SINR. \\
		$P_r$ &- &Received power \\
		$P_N$ &-117~dBm$^{\dagger1}$ &Noise power \\
		$P_I$ &-95.4~dBm$^{\dagger2}$&Interference power\\
		$P_t$ &- &RF transmit power \\
		$L_o$ &-26~dB$^{\dagger3}$&Constant path-loss factor\\
		$h$ &-$^{\dagger4}$ &Channel fading factor\\
		$r$ &- &Contact distance \\
		$\alpha$ &3.68 &Path-loss exponent \\
		$\eta$ &4.0~mW/mW &Power conversion factor \\
		$P_o$ &210~mW &Electronics power overhead \\
		$\epsilon$ &- &Total energy-consumption. \\
		$P_t^*$ &- &Optimal Transmit Power. \\ 
		$\gamma^*$ &- &Optimal SINR. \\
		$\epsilon^*$ &- &Minimum energy-consumption. \\ 
		$\gamma_o$ &- &Target SINR. \\
		$\epsilon_o$ &- &Threshold energy-consumption. \\
		$\Phi$ &- &IoT devices point process. \\
		$\Psi$ &- &Gateways point process. \\
		$\lambda$ &- &Intensity of the gateway point process.\\
		$\delta$ &- &Hard-core distance.\\
		$f_{\Psi}$ &- &Contact distance pdf. \\
		$f_{P_I}$ &- &Interference power pdf. \\
		$f_{h}$ &- &Channel fading factor pdf. \\
		\hline 
	\end{tabularx}
	\begin{tabularx}{\textwidth}{ l}
		$^{\dagger1}$ Calculated from noise figure of 6 dB and bandwidth 125 kHz.\\
		$^{\dagger2}$ The average interference power based on the highly dense\\ populated area in Melbourne \cite{7194048}.\\
		$^{\dagger3}$ Winner II scenario A1 \cite{Cite_WinnerII}.\\
		$^{\dagger4}$ Based on a Rayleigh Distribution with an average of unity.\\
	\end{tabularx}
	\label{Table_Notations}
\end{table}
\begin{equation}\label{Eqn_Shannon}
T_m = \frac{N_m}{B \log_2(1+\gamma)},
\end{equation} 
where $B$ is the bandwidth of the channel and $\gamma$ is the SINR seen by the gateway and is given by,
\begin{equation}\label{Eqn_SINR_Basic}
\gamma = \frac{P_r}{P_I+P_N},
\end{equation}
where $P_I$ is a random variable representing the interference power, $P_N$ is the noise power measured, and $P_r$ is the received power which can be calculated by,
\begin{equation}\label{Eqn_PathLoss}
P_r = L P_t,
\end{equation}
where $P_t$ is the transmit power of the end-device, and $L$ is the total loss factor between the end-device and its communicating gateway. The loss can be modeled for example by using the log-distance path-loss model with a fading component, i.e. $L={L_o} h r^{-\alpha} $ \cite{7386667}, where $L_o$ is a constant determined by the antenna gain, frequency, and propagation environment. $h$ is a random variable that represents the fading in the channel. $\alpha$ is the path-loss exponent and $r$ is the contact distance defined as the distance between the end-device and its gateway which is also a random variable and its statistics rely on the network regularity. The electric power consumption can be correlated with the RF power using a simple linear relation which follows the linear region in the power amplifier of the transmitter where this assumption is valid as per our measurements present in the Appendix section. Thus the total energy consumed by transmitting a fixed length message is given by,
\begin{equation}\label{Eqn_Eta}
\epsilon = (\eta P_t + P_o) T_m,
\end{equation}
where $\eta$ is the conversion factor of the power amplifier from \textit{electric} power to \textit{RF} power and $P_o$ is the electronic power consumption overhead incurred in the communication module to encode a message \cite{7386667}. By substituting the above relationships, a compact system model for the energy-consumption of the end-device is represented by,
\begin{equation}\label{eq_energy_main}
\epsilon = \left[ \frac{\eta}{L}\left(P_I+P_N\right) \gamma+ P_o\right] \frac{N_m}{B \log_2(1+\gamma)}.
\end{equation}
From the above relation, we note that the target SINR $\gamma$ is a free variable that the system can control by varying the transmit power whereas all the other parameters are related to hardware and channel limitations. This implicitly assumes that the IoT end-device is capable of adjusting its transmit power to achieve a certain average target SINR. It is worthwhile to note that the provided equation in (\ref{eq_energy_main}) can still accommodate access technologies that embed the control and payload messages in the same radio frame. In some cases where more complex access systems are utilized, heavy exchange of signaling might be required between the gateway and end-devices. In such cases, this proposed model has limited ability to capture the possible frame losses in the control plane and thus the energy required to re-transmit the signaling frames. However, the aim of this work is to characterize the lowest possible energy bound irrespective of the specific implementation of the access system. Nevertheless, many IoT applications are based on stationary devices having a steady radio channel loss and heavy signaling is not expected to occur frequently during the life span of the device in efficient access LPWAN systems.

\section{Link-Level Energy-Consumption Bound}\label{Sec_EnerBound}
\begin{figure}
	\normalsize
	\centering
	\includegraphics[width=\linewidth]{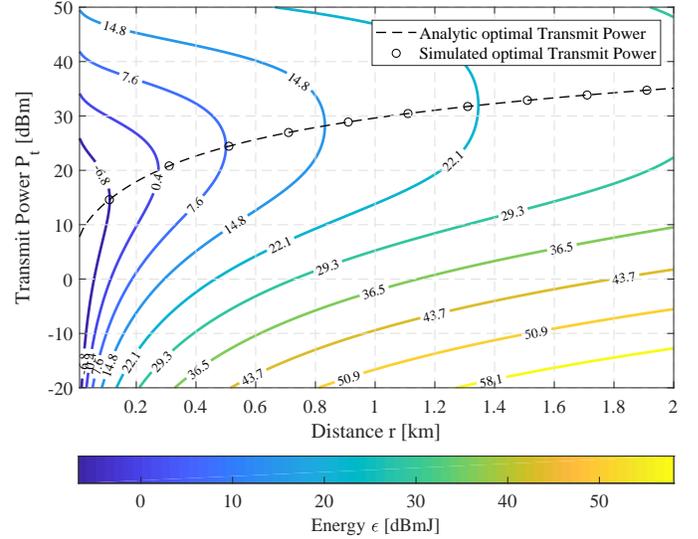}
	\caption{Contour demonstrating the effect of varying the distance between the end-device and its serving gateway and the transmit power on the energy-consumption levels based on (\ref{eq_energy_main}). The used parameters are listed in Table~\ref{Table_Notations}}
	\label{Fig_SNRDist}
\end{figure}

The energy-consumption model of the end-device is further analyzed to cover two scenarios; (i) the theoretical minimal energy-consumption boundary achieved at an optimal transmit power referred to as the unrestricted QoS scenario and (ii) a threshold energy-consumption model that is based on a target system SINR and is predefined by the designed application to achieve lower Message-Error-Rates (MERs)  and referred to as the restricted QoS scenario. Fig.~\ref{Fig_SNRDist} demonstrates the contour line  of the energy-consumption at the end-device by varying the transmit power and contact distance. To achieve a certain energy-consumption level, two different transmit power levels exist for a given contact distance due to varying the time-on-air. In most cases, the time-on-air is either restricted by the communication authorities or is limited by the Modulation and Coding Scheme (MCS) used.

\subsection{Unrestricted QoS Scenario}
In order to calculate the minimum bound, we search for the optimal transmit power that minimizes the overall energy-consumption of the communication module. This bound serves as a benchmark to the energy-consumption minimum of an end-device in a network. The equation (\ref{eq_energy_main}) is concave in nature and is differentiated in terms of the transmit power and equated to zero to search for the optimal RF transmit power as shown,
\begin{align}
P_t^* &= \argmin_{P_t} \left[ \epsilon \right] \nonumber \\
&=\text{Solution~of} \left[ \frac{\mathrm{d}}{\mathrm{d}P_t} \frac{A_1 \gamma+A_2}{\log_2\left(1+\gamma\right)} =0\right],
\end{align} \label{Eq_6}
where,
\begin{equation}\label{Eqn_Optimal_SNR_para}
A_1= \frac{\eta N_m}{BL} (P_I+P_N), \text{~~~and~} A_2= \frac{N_m P_o}{B}.
\end{equation}
The solution can be written in a closed form as follows,
\begin{equation}\label{Eqn_Optimal_Pt}
P_t^* = \frac{P_I+P_N}{L}\left[\exp\left(1+W\left[\frac{A_2/A_1-1}{e}\right]\right)-1\right],
\end{equation}
where $W[z]e^{W[z]}=z$ is the Lambert-W function. As a result, the SINR at the minimum bound is calculated as follows,
\begin{equation}\label{Eqn_Optimal_SNR}
\gamma^* = \exp\left(1+W\left[\frac{A_2/A_1-1}{e}\right]\right)-1,
\end{equation}
Accordingly, the minimum required electric energy-consumption for delivering $N_m$ bits is further given by,
\begin{equation}\label{Eq_Estart}
\epsilon^* = A_1\ln(2)\exp\left(1+W\left[\frac{A_2/A_1-1}{e}\right]\right).
\end{equation}

\begin{figure*}
\normalsize
\centering
\includegraphics[width=0.8\linewidth]{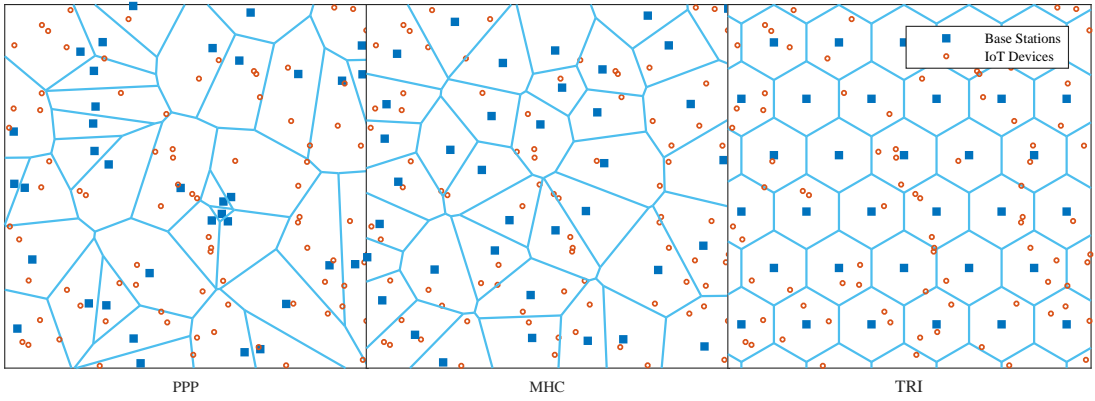}
\caption{The Voronoi tessellation of the three scenarios having equal gateway intensity, the regularity is increasing from left to right, PPP, MHC, and TRI.}
\label{Fig_Scenarios}
\end{figure*}

In Fig.~\ref{Fig_SNRDist}, the dashed line represents the analytic transmit power minimum in (\ref{Eqn_Optimal_Pt}) whereas the circles represent the simulated minimum transmit power. It is important to note that under this assumption, the presented optimal solution does not consider the effect of lengthening the transmission time on the quality of the service. Although in some applications, the optimal transmit power might be impractical due to low SINRs and ultimately high MERs, yet, this minimum bound is theoretically valid and can be used as a benchmark for energy optimality.

\subsection{Restricted QoS Scenario}
To maintain a certain level of quality, we restrict the transmit power from going below a certain threshold based on its corresponding system target SINR value. Considering some services that require tighter control on the delivery time of data, we assume that a fixed communication technology is used with a fixed MCS, thus a fixed data rate. In other words, for a given message size, the time-on-air is also constant. Under this assumption, the automatic power control mechanism for the IoT device will endeavor to maintain the target SINR. Naturally, the energy-consumption of the end-device is larger in magnitude than the minimum boundary and is given by,
\begin{equation}\label{Eq_restrictedQoS}
\epsilon_o = \left[ \frac{\eta}{L}\left(P_I+P_N\right) \gamma_o+ P_o\right] T_m ~ \ge \epsilon^*,
\end{equation}
where $\gamma_o$ is the target SINR required for the MCS to be reliably used \cite{8255054}. The optimum solution in (\ref{Eq_Estart}) will result in the lowest possible energy-consumption level since $\gamma_o\ge \gamma^*$. Moreover, in any practical system the throughput is lower than the channel capacity especially for finite-length messages and the time-on-air is also bounded as follows,
\begin{equation}
T_m \ge \frac{N_m}{B \log_2\left(1+\gamma_o\right)}.
\end{equation} 

\section{IoT Cellular Networks Energy-Consumption}\label{Sec_Cellular}
In a cellular network, IoT end-devices are connected to one or more gateways, aggregating the radio traffic and passing it through a back-haul link to the core. In some of those configurations, several gateways can serve one IoT device leading to redundancy, however it is not considered in this study. In order to obtain the energy-consumption bounds, an IoT device will opt to communicate with its serving gateway solely in order to minimize its power draw.

In order to establish a tractable model for device association, we assume that the devices are connected to their nearest gateways in terms of coverage distance. Thus the dominance area of a gateway is visualized by the Voronoi cell for every serving gateway. This assumption is valid when following a log-distance path-loss model or the well-known ITU cellular models~\cite{series2009guidelines}. Ideally, a cellular network would be deployed in a triangular lattice (forming a hexagonal shaped Voronoi tessellation), yet due to multiple practical reasons such as geography, high costs of deployment, and high capacity demands, a regular shape cannot be always maintained. Thus, a realistic deployment of a cellular network would be far from the perfect regularity of a lattice.

In this section, we demonstrate the differences in energy-consumption as a function of the network regularity. To quantify these differences, we study the two extremes in terms of regularity; a perfectly regular shaped network materialized in a trilingual lattice (TRI), and a fully irregular network based on Poisson point process (PPP). In addition, we study a third point process, Mat\'{e}rn Hard-Core point process (MHC)~\cite{7511676} which lies between these two extremes in terms of regularity. Fig.~\ref{Fig_Scenarios} depicts the three cases analyzed in this paper. IoT end-devices are depicted as an independent PPP since we have no control over their location. The end-devices' intensity is naturally larger due to the high number of devices expected to connect to the network. In this analysis, all point processes are depicted to be homogeneous. 

\subsection{Contact Distance and Network Regularity}
We model that end-devices as a PPP independent of the gateway point process. The distance between the IoT device and its nearest gateway is defined as the contact distance. The contact distance has a well-defined cumulative distribution function (cdf) conditional on the gateway point process $\Psi$ representing the statistics of all the Voronoi cells in a network. The contact distance cdf is well known to be~\cite{Book_Haenggi},
\begin{equation}\label{Eqn_CDFContact}
F_{\Psi}=\mathbb{P}(||o-\Psi|| \leq r)=\mathbb{P}(N(b(o,r))>0).
\end{equation} 
where $N(b(o,r))$ is the total number of points in set $\Psi$ located inside a ball with radius $r$. Since we have control over the gateway distribution, we analyze the three stochastic distributions:
\subsubsection{Gateways as a PPP}
The gateways are assumed to take a PPP distribution and the contact distance distribution derived from (\ref{Eqn_CDFContact}) is shown as~\cite{Book_Haenggi},
\begin{equation}\label{Eqn_PPP_CDF}
F_{PPP}=1-\exp(-\lambda \pi r^2),
\end{equation}
and the probability density function (pdf) as~\cite{Book_Haenggi},
\begin{equation}\label{Eqn_PPP_PDF}
f_{PPP}=2\lambda \pi r \exp(-\lambda \pi r^2).
\end{equation}
where both distributions are dependent on the intensity $\lambda$ of the point process.
\subsubsection{Gateways as a MHC}
The gateways are modeled to take a distribution of a Mat\'{e}rn hard-core process type-II~\cite{Book_Haenggi} which has a lower entropy than PPP due to a predefined thinning process. An initial parent PPP with $\lambda_b$ undergoes a thinning transformation satisfying the hard-core condition resulting in limiting the gateways from coexisting within a conditioning hard-core distance $\delta$ thus limiting the randomness of the distribution and increasing regularity. The intensity of the initial PPP $\lambda_b$ is assumed to be infinity since it has little effect on the thinning process result to simplify the outcome. The resulting intensity of the MHC is shown as,
\begin{equation}\label{intensity}
\lambda = \lim_{\lambda_b \to \infty}\frac{1-\exp(-\lambda_b \pi \delta^2)}{\pi \delta^2} =\frac{1}{\pi \delta^2}.
\end{equation}
The hard-core thinning process is based on a conditional thinning probability $\xi$ results in the MHC contact distance CDF is derived as $F_{MHC}=1-\exp\left(-2\pi \int_{0}^{R}r\xi dr\right)$, where the conditional thinning probability $\xi$ is~\cite{7511676},
\begin{equation}\label{Eqn_MHC_CDF_2}
\xi = \frac{\lambda}{1 - \lambda l},
\end{equation}
where $l$ is the asymmetrical lens due to the intersection of the points of the parent PPP and the new MHC points is~\cite{7511676},
\begin{equation}\label{Eqn_MHC_CDF_3}
l =
\begin{cases}
\pi r^2, & r < \frac{1}{2\sqrt{\pi \lambda}}\\
r^2 \cos^{-1}(\frac{2\pi \lambda r^2-1}{2\pi \lambda r^2})+\frac{1}{\pi \lambda} \cos^{-1}( \frac{1}{2\sqrt{\pi\lambda}r})\\
- \frac{1}{2\sqrt{\pi\lambda}} \sqrt{4r^2-\frac{1}{\pi\lambda}}, & r \geq \frac{1}{2\sqrt{\pi \lambda}}
\end{cases}
\end{equation}
and its pdf is $f_{MHC}=2\pi r \xi\exp\left(-2\pi\int_{0}^{R}r\xi dr\right)$. Since the pdf cannot be simplified analytically due to mathematical limitations, it is calculated numerically.
\begin{figure}
	\centering
	\includegraphics[width=\linewidth]{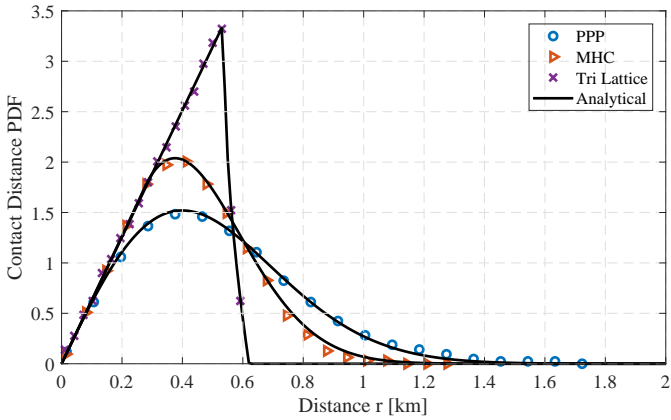}
	\caption{Contact distance pdf comparison between the PPP, MHC, and TRI cases respectively with $\lambda=1.0$ [gateways/km$^2$].}
	\label{Fig_CDF_Plot}
\end{figure}
\subsubsection{Gateways as a triangular lattice}
A triangular lattice is represented by regular hexagons and is deterministic in its shape. Although gateways are practically dispersed in a more irregular manner, analyzing a triangular lattice network can provide insight into the optimal dispersion and is set as a benchmark for coverage. The triangular lattice has an intensity $\lambda$ which can be derived from the lattice side length, $s = (2/\sqrt{3} \lambda)^{1/2}$. The contact distance cdf is shown in a closed form as follows~\cite{tri},
\begin{equation}\label{Eqn_TRI_CDF}
F_{TRI}=
\begin{cases}
\pi\lambda r^2, & r \leq \sqrt{\frac{1}{2\sqrt{3}\lambda}} \\
\pi\lambda r^2 + \sqrt{6\sqrt{3}\lambda r^2-3}\\
-6\lambda r^2 \cos^{-1}(\sqrt{\frac{1}{2\sqrt{3}\lambda r^2}}), & \sqrt{\frac{1}{2\sqrt{3}\lambda}} < r \leq \sqrt{\frac{2}{3\sqrt{3}\lambda}}\\
1, & r > \sqrt{\frac{2}{3\sqrt{3}\lambda}}
\end{cases}
\end{equation}
and the pdf as follows~\cite{tri},
\begin{equation}\label{Eqn_TRI_PDF}
f_{TRI}=
\begin{cases}
2\pi\lambda r, & r \leq \sqrt{\frac{1}{2\sqrt{3}\lambda}} \\
2\pi\lambda r \\ - 12\lambda r\cos^{-1}(\sqrt{\frac{1}{2\sqrt{3}\lambda r^2}}), & \sqrt{\frac{1}{2\sqrt{3}\lambda}} < r \leq \sqrt{\frac{2}{3\sqrt{3}\lambda}}\\
0, &r > \sqrt{\frac{2}{3\sqrt{3}\lambda}}
\end{cases}
\end{equation}
In order to compare between the three distinctive distributions, the different point processes are simulated with an identical intensity as shown in Fig.~\ref{Fig_CDF_Plot}. The tail of the pdf of the contact distance extends as the regularity decreases based on the location of the IoT sensors in the Voronoi cells. A device can be very far from its nearest gateway in the PPP case, however as the network becomes more regular, the distances from the serving gateway start to decrease as shown in the MHC case. In the triangular lattice, the devices are bound to have a maximum contact distance equivalent to the radius of the Voronoi cell and thus the pdf instantly diminishes at that radius.

\subsection{Interference and Channel Fading Effects}
\begin{figure}
	\centering
	\includegraphics[width=0.9\linewidth]{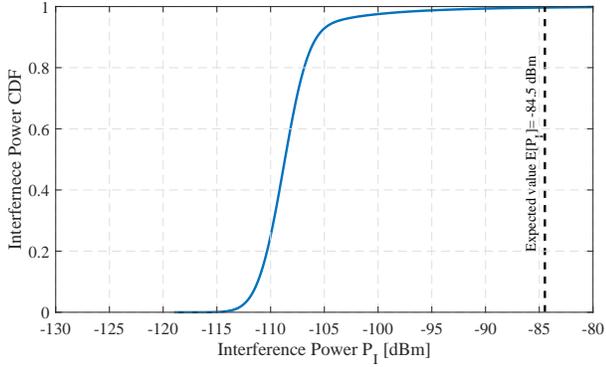}
	\caption{The interference statistics in the 915-928 MHz ISM-band as collected in the spectrum survey experiment, conducted in~\cite{7194048}. The dashed line corresponds to the average interference power.}
	\label{Fig_Interference}
\end{figure}
A random variable that also affects the energy-consumption is the instantaneous interference power in the spectrum denoted as $P_I$ in (\ref{eq_energy_main}). In this paper, we focus on the ISM-band which is part of the open access spectrum and supports IoT purposes among other technologies. The ISM-band is class-licensed, meaning that no subscription with telcom operators is required to access the spectrum. However, the downside is that many technologies will have to coexist in this band leading to high interference. Since our study focuses on a class-licensed frequency band under the assumption of highly congested traffic~\cite{8403749}, we solely consider the interference in the band regardless of its source. This approach is justified under the assumption that many networks will coexist in the spectrum reducing the control over the interference even if the optimal transmit power is utilized in one network, thus we focus the analysis on the end-device of interest instead. In some cases, spectrum authorities impose duty cycle restrictions to control the interference in the ISM-band, however due to the random access of the channel and class-licensed nature of the band, we need to consider the interference in the band regardless of its source. To derive the expected interference power, we capture the power spectral density (PSD) in the 915-928~MHz ISM-band~\cite{8403749} in Melbourne, Australia~\cite{7194048}. The statistics taken from that set of data of the RF power using 125 kHz measurement bandwidth are demonstrated in terms of the cdf as shown in Fig.~\ref{Fig_Interference}. The channel fading factor of the channel denoted as $h$ is a random variable has a Rayleigh distribution and $h\sim\exp(1)$.

\subsection{Cellular Energy-Consumption Analysis}
The energy-consumption model derived in (\ref{eq_energy_main}) is composed of three random processes; (i) the contact distance of the point process, (ii) the interference power of the spectrum and (iii) the fading factor of the channel. Thus, the energy-consumption model should take into consideration the statistics of those three random variables to calculate the expected energy-consumption $\mathbb{E}[\epsilon]$ for both the unrestricted and the restricted QoS scenarios derived previously.
\subsubsection{Unrestricted QoS Scenario}
The mean of the minimum energy-consumption based on the optimal RF transmit power in (\ref{Eq_Estart}) is given as follows,
\begin{equation}\label{Eq_ExpUnrest}
\mathbb{E}[\epsilon^*] = \int\epsilon^* f_{\Psi} f_h f_{P_I} dr dh dP_I
\end{equation}
\begin{figure}
	\centering
	\includegraphics[width=0.9\linewidth]{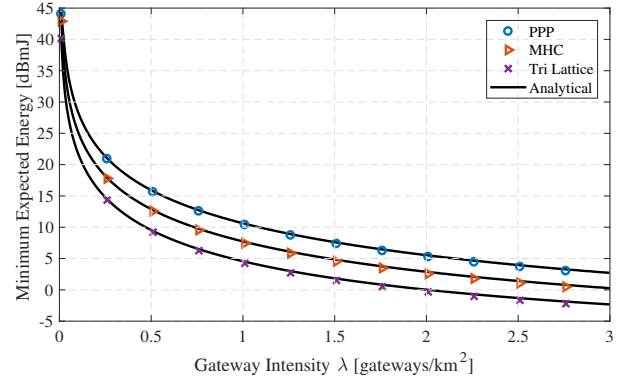}
	\caption{The expected minimum energy as a function of the gateway intensity for unrestricted QoS scenario.}
	\label{Fig_ExOptEn_vs_Lambda}
\end{figure}
\begin{figure*}
	\begin{equation}\label{Eqn_Expect_CDP}
	\mathbb{E}[r^\alpha]=\frac{\Gamma(\frac{\alpha+2}{2})}{(\pi \lambda)^{\alpha/2}}, \text{for PPP}
	\end{equation}
	\begin{equation}\label{Eqn_Expect_CDM}
	\mathbb{E}[r^\alpha]=2\pi\int_{0}^{R}r^{\alpha+1} \xi \exp\left(-2\pi\int_{0}^{R}r \xi dr\right) dr, \text{for MHC}
	\end{equation}
	\begin{equation}\label{Eqn_Expect_CDT}
	\mathbb{E}[r^\alpha] =\frac{1}{(\sqrt{3}\lambda)^{\alpha/2}}\frac{2^{\alpha/2}}{2+\alpha}\left(\frac{\sqrt{3\pi}}{2^\alpha}\frac{\Gamma(-\frac{(\alpha+1)}{2})}{\Gamma(-\frac{\alpha}{2})} +\left(\frac{1}{\sqrt{3}}\right)^\alpha\frac{4}{1+\alpha}F_{12}\left[\frac{1}{2},-\frac{(\alpha+1)}{2},\frac{1-\alpha}{2},\frac{3}{4}\right]\right), \text{for TRI}
	\end{equation}
\end{figure*}
where $f_{\Psi}$ is the point process pdf, $f_{P_I}$ is the interference power pdf, and $f_{h}$ is the channel fading coefficient pdf. This integral is performed numerically since the pdf of the interference power cannot be presented in a close form due to its spatio-temporal dependence. We simulate different intensity values for all three point processes and search for the minimum expected energy which is plotted versus the numerical integration in (\ref{Eq_ExpUnrest}) as shown in Fig. \ref{Fig_ExOptEn_vs_Lambda}.
As the regularity of the network increases, the expected minimum energy-consumption reduces regardless of the intensity. Naturally, as we increase the intensity of the gateways, the expected minimum energy decreases due to the contact distance distribution becoming very small.
\begin{figure}
	\centering
	\includegraphics[width=\linewidth]{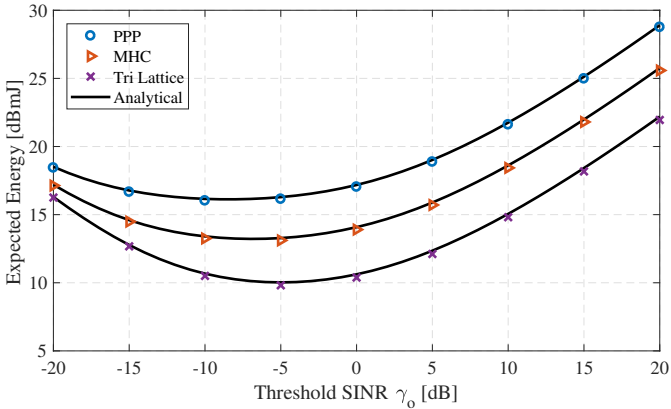}
	\caption{Expected energy model as a function of SINR threshold for each point process with intensity $\lambda=0.5$ [gateways/km$^2$].}
	\label{Fig_CellUnrePlot}
\end{figure}
\subsubsection{Restricted QoS Scenario}
\begin{figure}[h]
	\centering
	\includegraphics[width=0.9\linewidth]{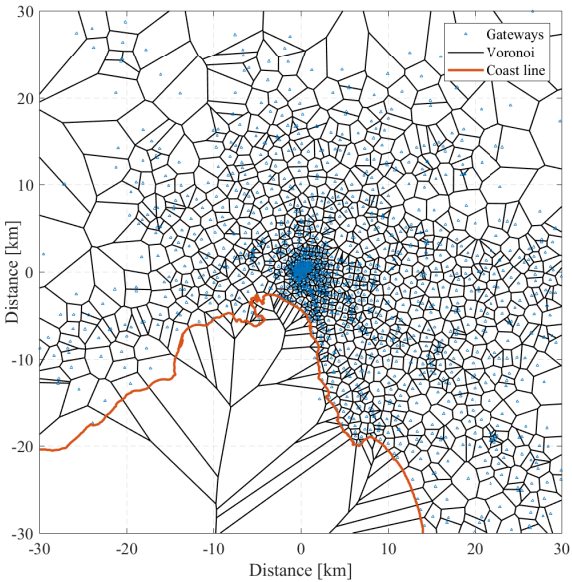}
	\caption{Realistic cellular network deployment of base stations in Melbourne, Australia for Telstra. Data is publicly available from ACMA \cite{Map}.}
	\label{Fig_Telstra}
\end{figure}
The target SINR is usually expected to be above a certain threshold to ensure a certain QoS keeping the MERs at a predefined value. Moreover, the maximum time-on-air is sometimes restricted by spectrum authorities. Both those factors lead to the minimum expected energy-consumption boundary being inapplicable in most cases. Therefore, we derive a more practical analytic equation in terms of the expected energy. The random variables are approximated to be independent assuming the network coexists with other networks and the number of users is very large (the average interference is not likely to change). The expected energy-consumption model for the QoS scenario as shown,
\begin{equation}\label{Eqn_Expect_Energy_QoS}
\mathbb{E}[\epsilon]=\left[ \frac{\eta\mathbb{E}[r^{\alpha}]}{L_o}\mathbb{E}[1/h]\left(\mathbb{E}[P_I]+P_N\right) \gamma_o+ P_o\right] T_m,
\end{equation}
To calculate the expected contact distance, we use $\mathbb{E}[r^\alpha]=\int_{0}^{R}r^{\alpha}f(r)dr$ based on the path-loss model, where $f(r)$ is the pdf of the contact distance which is based on the gateway point process. For gateways as a PPP case, the expected contact distance can be further simplified to the form shown in (\ref{Eqn_Expect_CDP}), where $\Gamma(.)$ is the complete gamma function. Similarly, in the MHC case, the expected contact distance is shown in (\ref{Eqn_Expect_CDM}). Finally, in the triangular lattice case, the expected contact distance is shown in (\ref{Eqn_Expect_CDT}), where $F_{12}[.,.,.,.]$ is the generalized hyper-geometric function. Fig.~\ref{Fig_CellUnrePlot} sheds light on the restricted QoS expected energy-consumption level where any chosen target SINR $\gamma_o$ yields an expected energy-consumption level of $\epsilon_o$ given a point process intensity $\lambda$.
\begin{figure}
	\centering
	\includegraphics[width=\linewidth]{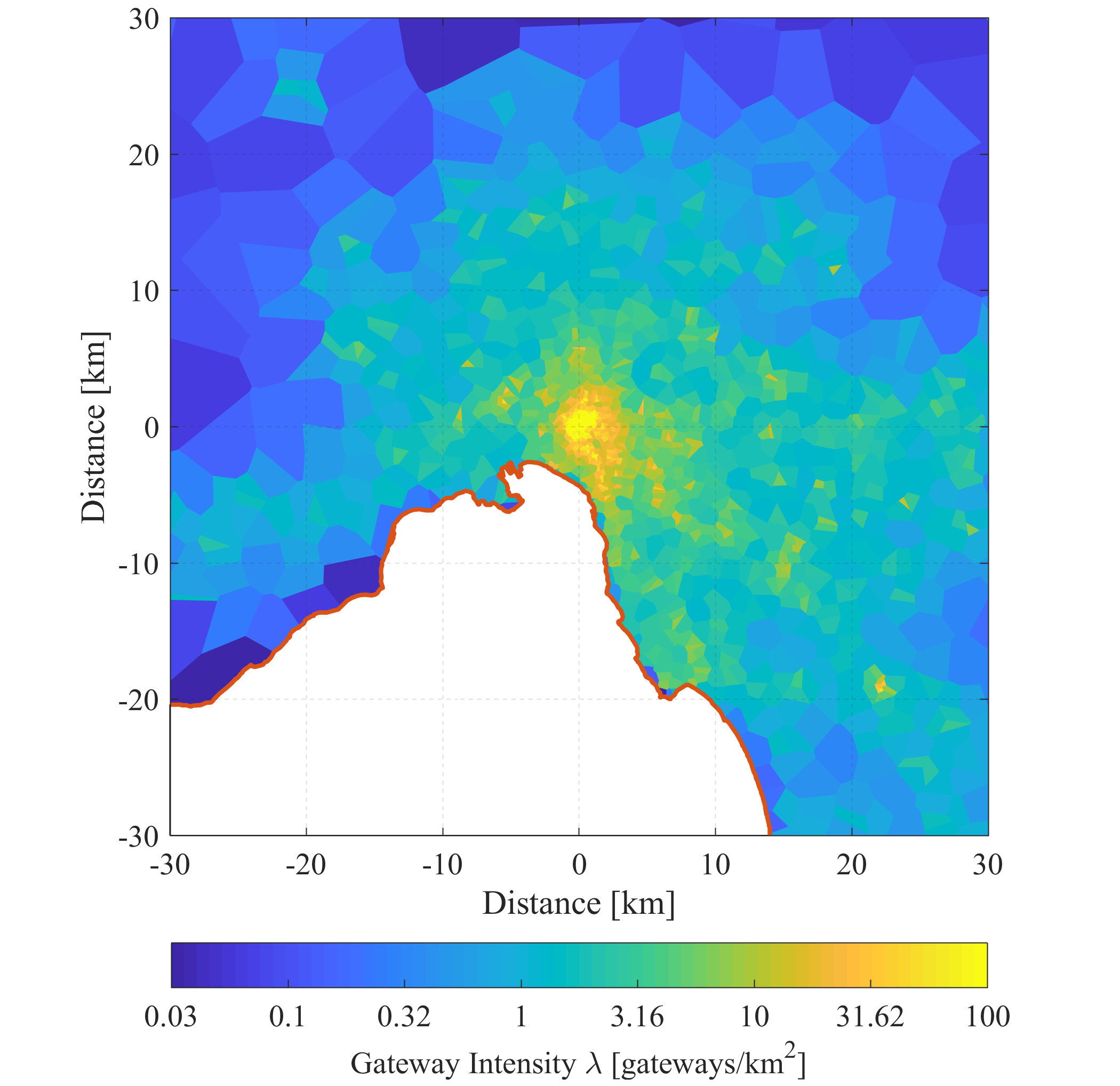}
	\caption{The individual density values of each Voronoi cell calculated as per (\ref{Eqn_lambda}) showing the transition in density from rural regions towards dense urban regions for Telstra scenario.}
	\label{Fig_lambdacont}
\end{figure}
\section{Energy-Consumption Bound for a Practical Network Deployment}\label{Sec_Operator}
\begin{figure}
	\centering
	\includegraphics[width=\linewidth]{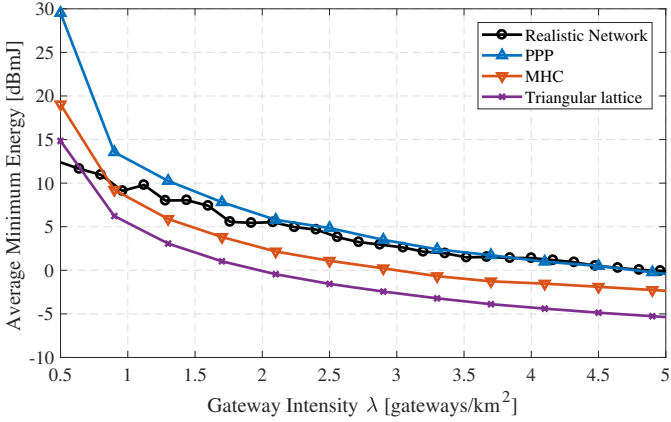}
	\caption{The expected minimum energy of realistic network deployment compared with the three theoretical geometric models (PPP, MHC, and triangular lattice).}
	\label{Fig_EnergyOperator}
\end{figure}

We apply the proposed energy-consumption framework on realistic network deployments consisting of three cellular operators in Australia; Telstra, Optus, and Vodafone. The locations of cellular base stations are obtained from ACMA public database \cite{Map}. As an example, Telstra network is shown in Fig.~\ref{Fig_Telstra} where the base stations are depicted along with their respective Voronoi cells. We assume that the base stations (BS) are acting as IoT gateways. We note that the intensity
of the base stations is inhomogeneous since it depends on the underlaying population density.
Since the intensity of BS is inhomogeneous across the map, we calculate the intensity of each Voronoi cell independently. Since each cell contains only one base station, the density is calculated as follows,
\begin{equation}\label{Eqn_lambda}
\lambda_i = \frac{1}{\mathcal{A}[C_i]},
\end{equation}
where $\mathcal{A}[C_i]$ is the area of the i-th Voronoi cell $C_i$. Fig.~\ref{Fig_lambdacont} represents the intensity variation across the map. In order to calculate the energy-consumption, we simulate an independent PPP to resemble the IoT end-devices. The contact distance and the minimum energy-consumption for each device are obtained based on (\ref{Eq_Estart}). The results are concatenated together. We take the average of the energy-consumption according to its corresponding intensity. We compare the resulting energy-consumption with the three theoretical point processes as depicted in Fig.~\ref{Fig_EnergyOperator}.
We note that the realistic performance of the energy-consumption converges to the PPP bound as the intensity of BS increases. The change in the BS regularity is due to network operators' interest in providing better coverage in rural areas whereas providing higher capacity in highly populated inner suburbs.

\section{Conclusion}\label{Sec_Conc}
\begin{figure}
	\centering
	\includegraphics[width=\linewidth]{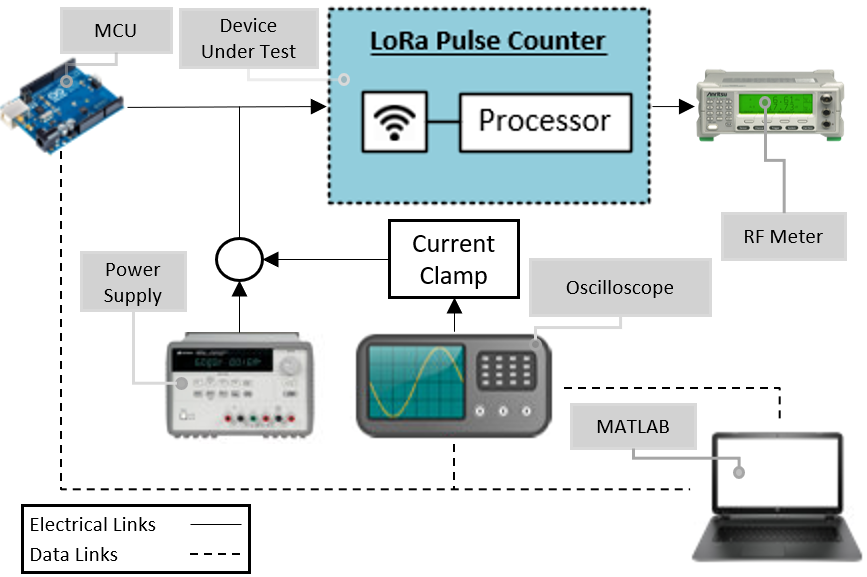}
	\caption{Schematic diagram for the measurement of electric power versus RF transmit power of the LoRa module.}
	\label{Fig_Diagram}
\end{figure}

\begin{figure}
	\centering
	\includegraphics[width=\linewidth]{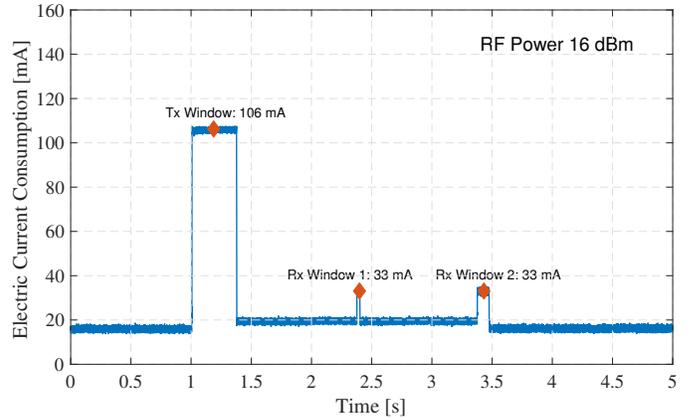}
	\caption{The electric current consumption of a typical transmission in LoRaWAN technology. Showing the transmit window and two receive windows. Measurements were performed on a LoRaWAN mDot module \texttrademark \cite{mDot} .}\label{Fig_Waveform}
\end{figure}

In this paper, we proposed an analytic framework to capture the lower bound of energy-consumption in cellular topologies for IoT applications as a benchmark for optimization purposes. The paper formulates the effects of network regularity on the energy-consumption bound using three geometrical models; PPP, MHC, and triangular lattice based on two scenarios: (i) unrestricted QoS scenario corresponding to the lowest possible bound and (ii) restricted QoS scenario where a target design system SINR exists. We compare the energy-consumption bounds with realistic network setups that exist in Australia. The average performance of realistic deployments is found to be bounded between the two geometrical extremes. Furthermore, the performance appears to follow the lower bound when the density is low to accommodate for coverage, while it converges towards the PPP model as the density of BS increases to accommodate for capacity. Our future work will involve optimizing the energy-consumption for practical IoT networks.
\appendix
We choose to present the measurements obtained from a LoRaWAN transceiver due to its simple protocol and clear power consumption profile. The test is done on the commercially-available mDot\texttrademark communication module~\cite{mDot} as depicted in the block diagram in Fig.~\ref{Fig_Diagram}. The module is powered by a regulated power supply, where the supply line is tapped through a current clamp having a fixed number of windings. The current clamp sensor converts Ampere measurements into a voltage signal without affecting the load on the power supply. The measurements are read using an oscilloscope which in turn stores them into MATLAB. Another MATLAB script is talking to a low level micro-controller unit (MCU) to control the module to send a fixed length message. Moreover, an RF power meter is connected to the module antenna output in order to measure the RF power.
The resulting instantaneous power-consumption measurement is shown in Fig.~\ref{Fig_Waveform} for one RF transmit power. This resembles a typical LoRaWAN Class A~\cite{7815384} device behavior, where a transmit window occurs followed by two receive windows. The message length is set to 18 bytes generated as a standard \textit{``join-request''} command \cite{specification2015lora}. We use the electric current measurement in the transmit window to calculate the corresponding electric power consumption knowing that the supply voltage is set to $3.6$~V.

\begin{figure}
	\centering
	\includegraphics[width=\linewidth]{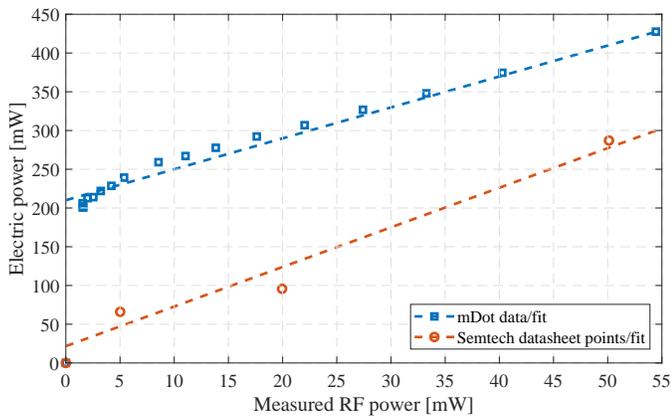}
	\caption{The measured current draw of the module during transmission, for different RF power settings. In comparison with the values from the datasheet of SEMTECH transceiver chipset~\cite{Semtech_chipset} without the controller over-head.}
	\label{Fig_eta}
\end{figure}

To obtain the conversion factor from electrical to RF power $\eta$ and the overhead power $P_o$ present in our derivation in (\ref{Eqn_Eta}), we measure the power consumption over different settings of the RF power during the transmission windows. By calculating the electric power of the transmit window for every RF power setting, we can estimate the conversion factor as depicted in Fig.~\ref{Fig_eta}. The electric power consumption can be very-well approximated to a linear relation with the actual RF power corresponding to the linear region in a typical power amplifier\footnote{The antenna transmit power is a $50~\Omega$ measured using a matched-load RF power-meter at the terminal of the module.}. The figure shows that the integrated communication module (the blue line) has a higher overhead than a typical stand-alone transmitter chipset~\cite{Semtech_chipset}. The mDot power readings are taken by experimental measurements whereas the chipset values are taken from the data sheet.

\ifCLASSOPTIONcaptionsoff
\newpage
\fi

\bibliography{Energy_Bounds}

\end{document}